\documentclass[10pt,a4paper]{article}
\usepackage[latin1]{inputenc}
\usepackage{amsmath}
\usepackage{amsfonts}
\usepackage{amssymb}
\author{H.F. Westman\footnote{\texttt{hwestman@physics.usyd.edu.au}} \, and T.G. Zlosnik\footnote{\texttt{tzlosnik@perimeterinstitute.ca}}}

\title{Gravitation and spatial conformal invariance}

\begin{document}
\maketitle

\begin{abstract}
It is well-known that General Relativity with positive cosmological
constant can be formulated as a gauge theory with a broken $SO(1,4)$
symmetry. This symmetry is broken by the presence of an internal
space-like vector $V^A$, $A=0,...,4$, with $SO(1,3)$ as a residual
invariance group. Attempts to ascribe dynamics to the field $V^{A}$ 
have been made in the literature but so far with limited
success. Regardless of this issue we can take the view that $V^A$
might actually vary across spacetime and in particular become null or
time-like. In this paper we will study the case where $V^A$ is null. This is shown
to correspond to a Lorentz violating modified theory of gravity.
Using the isomorphism between the de Sitter group and the
spatial conformal group, $SO(1,4)\simeq C(3)$, we show that the
resulting gravitational field equations are invariant under all the symmetries,
but spatial translations, of the conformal group $C(3)$.
\end{abstract}

\section{Introduction}
\label{intro}

General Relativity is often formulated using a metric tensor $g_{\mu\nu}$ and a metric-compatible and torsion free connection $\nabla_\mu$. However, a more powerful formulation (especially when it comes to coupling to spin-$\frac12$ matter fields) employs a co-tetrad $e^I_\mu$ and an $\mathfrak{so}(1,3)$-valued spin connection $\omega_\mu^{\ IJ}$ as independent variables\footnote{We will use the convention that differential forms will be introduced with spacetime explicit indices but thereafter referred to with those indices implicit e.g $e^{J}_{\mu}$ and $e^{J}\equiv e^{J}_{\mu}dx^{\mu}$}. The spacetime metric $g_{\mu\nu}$ is recovered using the relation 
\begin{eqnarray}
g_{\mu\nu} = \eta_{IJ} e^{I}_{\mu}e^{J}_{\nu}
\end{eqnarray}
where $\eta_{IJ}= diag(-1,1,1,1)$. In the language of differential forms \cite{Frankel:1997ec}, the gravitational action takes on a particularly simply form referred to as the Palatini action:
\begin{eqnarray}
\label{spala}
{\cal S}_{P}(e^{I},\omega^{IJ})=  \kappa_{P} \int \epsilon_{IJKL}\left(e^{I}\wedge  e^{J} \wedge R^{KL}-\frac{\Lambda}{6} e^{I}\wedge  e^{J} \wedge e^{K}\wedge e^{L}\right)
\end{eqnarray}
where $\epsilon_{IJKL}$ is the four-dimensional Levi-Civita symbol, $R^{I}_{\phantom{I}J} \equiv d \omega^{I}_{\phantom{I}J} + \omega^{I}_{\phantom{I}K} \wedge \omega^{K}_{\phantom{K}J}$ the curvature of $ \omega^{I}_{\phantom{I}J}$, $\Lambda$ the cosmological constant, and $\kappa_P$ the gravitational constant. The Palatini action reproduces exactly General Relativity in the absence of fermionic fields and is slightly modified in their presence by a non-zero torsion field $T^I \equiv de^I+\omega^I_{\ J}\wedge e^J\neq0$.

In this first order formulation the spin connection $\omega^{IJ}$ is naturally thought of as a standard gauge connection related to local Lorentz symmetry. However, it was realized by MacDowell and Mansouri \cite{MacDowell:1977jt} (though building on ideas by Cartan \cite{SharpeCartan}) that the natural group to use for General Relativity (with {\em positive} cosmological constant) is not $SO(1,3)$ but $SO(1,4)$, i.e. the group of transformations that preserve $\eta_{AB}=diag(-1,1,1,1,1)$ where $A,B=0,1,2,3,4$.
As opposed to the six parameters of the group $SO(1,3)$, the group $SO(1,4)$ has ten parameters; the connection coefficient one-forms $A^{\phantom{\mu}AB}_{\mu}$ may be decomposed as follows:  $h_{\mu}^{\phantom{\mu}I} \equiv A_{\mu}^{\phantom{\mu}4I}$ ,  $w_{\mu}^{\phantom{\mu}IJ} \equiv A_{\mu}^{\phantom{\mu}IJ}$ where recall that indices $I$ and $J$ range from $0,1,2,3$. Consequently the representation independent curvature 
$F^{\phantom{\mu\nu}AB}_{\mu\nu}$ may be decomposed as follows:

\begin{eqnarray}
\label{fijmm} F^{IJ}&=& d w^{IJ}+w^{I}_{K}\wedge w^{KJ}- h^{I}\wedge h^{J} \\
\label{sttor} F^{4I} &=& dh^{I}+ h_{J}\wedge w^{JI}
\end{eqnarray}
A more familiar interpretation of these differential forms is possible: We firstly identify $w^{IJ}$ as corresponding to the $SO(1,3)$ gauge field $\omega^{IJ}$. The field $h^{I}$, in simply being a part of $A^{AB}$, must have dimensions of inverse length if the prior identification of $w^{IJ}$ is appropriate. We suggestively identify $h^{I}$ with the co-tetrad $e^{I}$ divided by a constant length scale $\ell$ i.e. $h^{I}= \frac{1}{\ell}e^{I}$. Therefore we have that:

\begin{eqnarray}
F^{IJ}=R^{IJ}-\frac{1}{\ell^2}e^I\wedge e^J\qquad  F^{4I}=\frac{1}{\ell}T^I.
\end{eqnarray}
where again $T^I$ is the torsion. Therefore $F^{AB}$ may be decomposed into tensors familiar from the $SO(1,3)$ perspective.

To better understand what is going on here from a geometric point of view we visualize the spacetime manifold as embedded in a five-dimensional space. On top of any  point on the manifold we imagine a symmetric space called a {\em model space}, which in this case is a de Sitter spacetime\footnote{A de Sitter spacetime is the four-dimensional manifold defined by $X^A X^B\eta_{AB}=\ell^2$.}, and exhibits the symmetry $SO(1,4)$. For each model space one point is singled out as the contact point of the model space and the four dimensional manifold. Consider now what happens when we apply an $SO(1,4)$ symmetry transformation. A subgroup of those transformations will be such that the contact point is left invariant while others will change the point of contact on the model space. Those which leave the point of contact invariant we identify as local Lorentz transformations $SO(1,3)$ and the corresponding gauge connection one-form is $\omega^{IJ}$. Those transformations that change the point of contact we identify as translations (or more accurately {\em transvections} as we are dealing with a de Sitter model space and not a flat manifold), and the corresponding gauge connection one-form is $h^I$. Thus, we understand that the co-tetrad can be viewed as a gauge connection related to local translations, i.e. spacetime diffeomorphisms. It should be stressed that the relationship between local translations, generated by a contact point changing $SO(1,4)$
transformations, and a diffeomorphism, is perhaps not straightforward and has been the source of some disagreement \cite{Petti:2006ue}.

The action
\begin{eqnarray}
\nonumber {\cal S}_{MM}&=&\kappa_{MM}\int  \epsilon_{ABCD4}F^{AB}\wedge F^{CD}=\kappa_{MM}\int  \epsilon_{IJKL}F^{IJ}\wedge F^{KL}  \\
\label{ECMM}  &=& -2\kappa_{MM}\int  \epsilon_{IJKL}\left(\frac{1}{l^{2}}e^{I}\wedge e^{J}\wedge R^{KL}- \frac{1}{2l^{4}}e^{I}\wedge e^{J}\wedge e^{K} \wedge e^{L}
  - \frac{1}{2}R^{IJ} \wedge R^{KL}\right) \label{mmm1}
\end{eqnarray}
yields the Palatini action with a positive cosmological constant as well as  the boundary term $\epsilon_{IJKL}R^{IJ}\wedge R^{KL}$ 
\footnote{For a self-contained discussion of the route to General Relativity from actions such as (\ref{ECMM}) we refer the reader to the two very readable accounts \cite{Wise:2006sm}, \cite{Randono:2010cq}.}.  Comparison of (\ref{ECMM}) with the Palatini action (\ref{spala})  immediately gives the relations $\kappa_{P} = -2\kappa_{MM}/l^{2}$, $\Lambda= 3/l^{2}$. Thus we can think of $\ell$ as a
conversion factor between the cosmological de Sitter scale and the arbitrary length unit as defined by the meterstick in Paris which is ultimately determined by the physics of elementary particles.
However, from the current perspective it would be natural to work in units in which $\ell=1$ \footnote{Compare to the naturalness of using $c=1$ in conventional relativistic physics.} , or equivelently to directly identify $h^{I}$ with the co-tetrad $e^{I}$. In fact, we shall do so in this paper. In such units
$\Lambda=3$.

However, note that the action \eqref{ECMM} is \emph{not} invariant under $SO(1,4)$ gauge transformations due to the arbitrary selection of a `direction' indicated by the $4$ in the five dimensional Levi-Civita symbol. This signals the presence of a non-dynamical `absolute object' \cite{AndersonRelativity, Westman:2007yx} which can be made explicit by introducing the five-dimensional `space-like' vector $V^A$ (with $|V|^{2}\equiv V^AV^B\eta_{AB}=1$) which takes the form $V^A\overset{*}{=}(0,0,0,0,1)$ in a privileged $SO(1,4)$ gauge. Using the `absolute object' $V^A$ the action becomes
\begin{eqnarray}
\label{ECMM2}
{\cal S}_{MMV}  &=& \kappa_{MM}\int  \epsilon_{ABCDE}V^{E}F^{AB}\wedge F^{CD}.
\end{eqnarray}
In order to reproduce the Palatini action it is important that we regard $V^A$ as an {\em \`a priori} postulated object that should not be varied when extremizing the action. The presence of `absolute objects' (such as the Minkowski metric in special relativity) is often taken as an indication that the theory is incomplete. One way to proceed would be to introduce dynamics for the absolute object (which is what happened in the transition from special relativity with the {\em \`a priori} postulated non-dynamical Minkowski metric $\eta_{\mu\nu}$  to general relativity with a dynamical spacetime metric $g_{\mu\nu}$). From this perspective it is natural to provide some dynamics for $V^A$. In fact, it has proven very difficult to provide a kinetic term to $V^A$ \cite{Randono:2010cq,Randono:2010ym}. A method of enforcing the desired behaviour of $|V|^{2}$ would be to simply enforce a fixed normal condition upon $|V|^{2}$ via the addition of a Lagrangian constraint to the action \cite{Stelle:1979va}. We shall not prescribe precise dynamics for $V^{A}$, but shall assume that such dynamics may exist so that the field $V^{E}$ itself may be considered not merely a Lagrange multiplier field enforcing a constraint upon ${\cal S}_{MMV}$. It is with this in mind that we will refer to (\ref{ECMM2}) as an $SO(1,4)$ invariant action.

Given that in general $V^{A}$ might be expected to have some dynamics, it is conceivable that there exist regimes where the norm $|V|^{2}$ may be approximately equal to a value other than the space-like value which is known to lead to General Relativity. The purpose of this article is to examine in detail the behaviour of the `gravitational' action (\ref{ECMM2}) when $|V|^{2}$ is assumed to be \emph{null}. Towards these ends we initially take what may seem like a diversion in establishing the isomorphism between the group $SO(1,4)$ and the conformal group the three dimensional conformal group $C(3)$

\section{The isomorphism between $SO(1,4)$ and the conformal group $C(3)$}

Recall that the conformal group $C(m)$ may be defined as the set of all coordinate transformations $x^{m}\rightarrow \tilde{x}^{m}$ that leave invariant the m-dimensional Euclidean metric $\delta_{pq}$ up to a conformal factor dependent upon $\tilde{x}^{p}$. We represent this by a general $\tilde{x}^{m} = f(\theta_{\Omega},J^{\Omega},x^{n})$,  where the function $f$ is taken to depend upon the $\Omega=\frac{1}{2}(m+2)(m+1)$ generators $J^{\Omega}$ of $C(m)$ and associated small parameters $\theta_{\Omega}$.
The generators $J^{\Omega}$ may be classified by the effect on coordinates, expressed as infinitesimal transformations:  $m$ operators $P_{m}$ generate translation coordinate transformations $\tilde{x}^{m}= x^{m} + a^{m}$;  $\frac{1}{2}m(m-1)$ operators $s_{mn}$ which generate rotations $\tilde{x}^{m} = x^{m} + R^{m}_{\phantom{i}n}x^{n}$ ($R_{mn}=-R_{nm})$  ; one operator $D$ which generates a dilation $\tilde{x}^{m} = (1+\epsilon)x^{m}$ ; and $m$ operators $k_{m}$ that generate special conformal transformations $\tilde{x}^{m} =x^{m}+2x^{n}b_{n}x^{m}-x^{2}b^{m}$ ($x^{2}\equiv \delta_{mn}x^{m}x^{n}$).

These generators satisfy the following Lie algebra:

\begin{eqnarray}
\left[D,p_{m}\right] &=& i p_{m}\\
\left[D,k_{m}\right] &=& -i k_{m} \\
\left[k_{m},p_{n}\right] &=& 2i\left(\delta_{mn}D-s_{mn}\right) \\
\left[k_{m},s_{np}\right] &=& i\left(\delta_{mn}k_{p}-\delta_{mp}k_{n}\right)\\
\left[p_{m},s_{np}\right] &=& i\left(\delta_{mn}p_{p}-\delta_{mp}p_{n}\right) \\
\left[s_{mn},s_{pq}\right] &=& i\left(\delta_{np}s_{mq}+\delta_{mq}s_{np}-\delta_{mp}s_{nq}-\delta_{nq}s_{mp}\right)
\end{eqnarray}
where recall that labels run from 1 to $m$. We now focus exclusively on the case $m=3$.
This algebra may be written in a more compact form via the following definitions:

\begin{eqnarray}
\label{iso1}S_{np} &\equiv& s_{np} \\
S_{04} &\equiv & D \\
S_{0p} &\equiv & \frac{1}{2}\left(p_{p}-k_{p}\right) \\
\label{iso4} S_{4p} &\equiv & \frac{1}{2}\left(p_{p}+k_{p}\right)
\end{eqnarray}
Or, more compactly, using the notation $\eta_{AB}=diag(-1,1,1,1,1)$:

\begin{eqnarray}
\left[S_{AB},S_{CD}\right] &=& i\left(\eta_{BC}S_{AD}+\eta_{AC}S_{BD}-\eta_{AC}S_{BD}-\eta_{BD}S_{AC}\right)
\end{eqnarray}
Though this remains the Lie algebra for $C(3)$, we see that this is identical to the Lie algebra for the group $SO(1,4)$.
In this sense the equations (\ref{iso1})-(\ref{iso4}) codify the isomorphism between the groups $C(3)$ and $SO(1,4)$. 
Therefore a representation of $SO(1,4)$ is also a representation of $C(3)$, and by implication 
an $SO(1,4)$ invariant action such as $(\ref{ECMM2})$ is also invariant under $C(3)$ transformations. Therefore,
General Relativity with a positive cosmological constant may be seen as a theory of broken spatial conformal invariance. Recall that in the case of Macdowell-Mansouri gravity, the $SO(1,4)$ invariance was broken down to a residual $SO(1,3)$ invariance preserving the form $V^{A}=(0,0,0,0,1)$. We can see immediately from equations (\ref{iso1}) to (\ref{iso4}) that from the $C(3)$ perspective this form for $V^{A}$ is not invariant under dilations, or combined $(p_{p}+k_{p})$ transformations. The residual $SO(1,3)$ gauge invariance of General Relativity (with a positive cosmological constant) may be seen as invariance under the generators $s_{np}$ and $(p_{p}-k_{p})$ of the group $C(3)$.

However, as we shall see, there are alternative forms for $V^{A}$ which secure more natural residual gauge invariance from the $C(3)$ perspective. A motivating feature behind the Macdowell-Mansouri approach was to essentially be able to invoke behaviour  of an object such as $V^{A}$ to separate in the gauge field $A^{AB}$ the co-tetrad from the spin connection (hence the residual $SO(1,3)$ symmetry).

\section{The case where $V^{A}$ is null}

Under an $SO(1,4)$ (or equivalently, $C(3)$) transformation, the components of $V^{A}(x^{\mu})$ transform as follows:

\begin{eqnarray}
\label{vtran}
\tilde{V}^{C} =  (e^{-\frac{i}{2}\Theta^{AB}(x^{\mu})S_{AB}})^{C}_{\phantom{C}D}V^{D} \label{trans1}
\end{eqnarray}

A clear geometric interpretation of this is a rotation in a five dimensional internal space with metric $\eta_{AB}$. As such, it is always possible to locally choose `coordinates' such that a null vector $V^{A}$ has the following components \footnote{We will assume throughout that $\phi\neq 0$. The possibility $V^{A}=0$ must be considered separately.}:

\begin{eqnarray}
V^{4}&=&\phi(x^{\mu}) \qquad V^{0}= \pm \phi(x^{\mu}) \qquad V^{i}= 0 \label{co3}
\end{eqnarray}
where henceforth lowercase latin indices $i,j,k,l$ will vary from $1$ to $3$, labeling space-like directions in the \emph{internal} space. One may now ask what residual gauge invariance preserves the form (\ref{co3}). Using the transformation law (\ref{vtran}), the representation of the generators $S_{AB}$ given by (\ref{sp1reps}), and the relations (\ref{iso1}) to (\ref{iso4}), it is readily seen that the form (\ref{co3}) represents two possibilities. The case $V^{A}=(\phi,0,0,0,-\phi)$ is invariant under 
transformations generated by $s_{ij}$ and $k_{i}$, and transforms homogeneously under $D$. The case $V^{A}=(\phi,0,0,0,\phi)$ meanwhile is invariant under transformations generated by $s_{ij}$, $p_{i}$,  and transforms homogeneously under $D$, but with opposite weight to the former case i.e. for a dilation associated with a small parameter $\delta$ we have $\tilde{V}^{A}\rightarrow (1-\delta)V^{A}$ for the case $V^{0}=-V^{4}$ and $\tilde{V}^{A}\rightarrow (1+\delta)V^{A}$ for the case $V^{0}=+V^{4}$.
Therefore the null $V^{A}$ case displays residual gauge invariance and covariance more cleanly interpreted in terms of $C(3)$ variables than in the space-like case. We may still visualize the spacetime manifold as embedded in a five-dimensional space. However, the functional form (\ref{co3}) now implies a preferred temporal direction in this space. The residual gauge freedom for the form (\ref{co3}) is indeed not $SO(1,3)$ invariance and so on the spacetime manifold violations of local Lorentz invariance are to be expected. We now expand the connection coefficient one-form $A^{AB}$ in terms of the conformal group generators:

\begin{eqnarray}
\label{adecomp1}A^{\phantom{\mu}C}_{\mu\phantom{C}D}  &=&     \mathbf{E}^{i}_{\mu} \left(p_{i}\right)^{C}_{\phantom{C}D} + \mathbf{F}^{i}_{\mu} \left(k_{i}\right)^{C}_{\phantom{C}D}  + \mathbf{C}_{\mu} \left(D\right)^{C}_{\phantom{C}D}  + \frac{1}{2} \mathbf{W}^{ij}_{\mu}\left(s_{ij}\right)^{C}_{\phantom{C}D} \\
\label{adecomp2}&=& h^{i}_{\mu}\left(S_{4i}\right)^{C}_{\phantom{C}D}+ h^{t}_{\mu}\left(S_{4t}\right)^{C}_{\phantom{C}D}
      + \mathbf{b}^{i}_{\mu}\left(S_{ti}\right)^{C}_{\phantom{C}D}+ \frac{1}{2} \mathbf{W}^{ij}_{\mu}\left(s_{ij}\right)^{C}_{\phantom{C}D} 
\end{eqnarray}
where we have introduced the spacetime fields $ \mathbf{E}^{i}_{\mu}$, $ \mathbf{F}^{i}_{\mu}$, $ \mathbf{C}_{\mu} $ and $ \mathbf{W}^{ij}_{\mu}$, the physical interpretation of which will be later apparent. Furthermore we decompose all fields into their time and spatial components:

\begin{eqnarray}
 \mathbf{E}^{i}_{\mu} &=& ({\cal E}^{i}, E^{i}_{m}) \qquad \mathbf{F}^{i}_{\mu} = ({\cal F}^{i}, F^{i}_{m})  \qquad
   \mathbf{C}_{\mu}= ({\cal C}, C_{m} ) \qquad \mathbf{W}^{ik} = (\Omega^{ik},\omega^{ik}_{m})
 \end{eqnarray}
 and
 
 \begin{eqnarray}
 h^{i}_{\mu}  &=& (\epsilon^{i}, h^{i}_{m})  \qquad  h^{t}_{\mu} = -\mathbf{C}_{\mu} \qquad  \mathbf{b}^{i}_{\mu} =  (\beta^{i}, b^{i}_{m}) 
  \end{eqnarray}
We will use the lowercase Latin letters $m,n,p,q$ exclusively to refer to spatial components on the spacetime manifold.
The action (\ref{ECMM2}) is constructed from three distinct quantities: $\epsilon_{ABCDE}$, $V^{A}$, and $F^{AB}$. We concentrate firstly on the former two quantities. 
We first assume that $V^{A}$ takes the functional form given by (\ref{co3}); to reflect this, we introduce the following `internal forms' $T_{A} \equiv \delta^{0}_{A}$ and $S_{A} \equiv \delta^{4}_{A}$, and hence: 

\begin{eqnarray}
V^{E}= \phi S^{E}\pm \phi T^{E}
\end{eqnarray}
where indices have been raised with $\eta^{AB}$, e.g. $T^{0}=-1$
Given the quantities $S_{E}$ and $T_{E}$ we may define a `spatial' Levi-Civita symbol $\varepsilon_{ABC}$ (which is non-vanishing only 
for indices values $1..3$) via the following relation:

\begin{eqnarray}
\epsilon_{ABCDE}= 5\cdot 4T_{[A}\varepsilon_{BCD}S_{E]}
\end{eqnarray}
The quantity that appears in the action (\ref{ECMM2}) is $\epsilon_{ABCDE}V^{E}$, and this may be developed as follows:

\begin{eqnarray}
\epsilon_{ABCDE}V^{E} &=& 5\cdot 4 T_{[A}\varepsilon_{BCD}S_{E]}
V^{E}  \\
   &=& 5\cdot 4  T_{[A}\varepsilon_{BCD}S_{E]}( \phi S^{E}\pm  \phi T^{E} ) \\
   &=& 4\phi T_{[A}\varepsilon_{BCD]}  \pm 4\phi S_{[A}\varepsilon_{BCD]}
   \end{eqnarray}
Explicitly, the action (\ref{ECMM2}) may then be written as:

\begin{eqnarray}
{\kappa}_{MM}\int\epsilon_{ABCDE}V^{E}F^{AB} \wedge F^{CD} &=& 4\kappa_{MM}\int\phi\left(T_{[A}\varepsilon_{BCD]}  \pm 4\phi S_{[A}\varepsilon_{BCD]}\right)F^{AB}\wedge F^{CD}
\end{eqnarray}

For simplicity, we now concentrate on the case where $V^{A}=(\phi,0,0,0,-\phi)$. As we shall see, this case has a rather clear interpretation.  Following this we shall compare this to the other case $V^{A}=(\phi,0,0,0,\phi)$. Due to the presence of a preferred temporal direction implied by (\ref{co3}),  we would like to decompose $A^{AB}$ and $F^{AB}$ into parts the contain time components and parts that do not. Then, to locally choose time to be the preferred time on the spacetime manifold should lead to a simplification of the equations of motion. 

Accordingly, we decompose $A^{AB}$ into temporal forms $\alpha^{AB}$ and spatial forms $a^{AB}$:  $A^{AB}=\alpha^{AB}+a^{AB}$ i.e. 

\begin{eqnarray}
\alpha^{AB} &=& A^{AB}_{t} dt \\
a^{AB} &=& A^{AB}_{m}dx^{m} 
\end{eqnarray}
Additionally we decompose the exterior derivative operator $d$ as follows: $d=d_t+^{(3)}\!d$ i.e.

\begin{eqnarray}
d_{t}\left(Y_{\mu\nu..} dx^{\mu}\wedge dx^{\nu} \wedge ..\right) &=&\left( \partial_{t}Y_{\mu\nu..} \right)dt\wedge dx^{\mu} \wedge dx^{\nu} \wedge .. \\
^{3}d\left(Y_{\mu\nu..} dx^{\mu}\wedge dx^{\nu} \wedge ..\right) &=&\left( \partial_{m}Y_{\mu\nu..} \right)dx^{m}\wedge dx^{\mu} \wedge dx^{\nu} \wedge .. 
\end{eqnarray}
Then we immediately obtain
\begin{eqnarray}
F^{AB}&=&dA^{AB}+A^{AC}\wedge A^{DB}\eta_{CD}\\
&=&(d_t+^{(3)}\!d)(\alpha^{AB}+a^{AB})+(\alpha^{AC}+a^{AC})\wedge(\alpha^{DB}+a^{DB})\eta_{CD}\\
&=&d_ta^{AB}+^{(3)}\!d\alpha^{AB}+^{(3)}\!da^{AB}+(\alpha^{AC}\wedge a^{DB}+a^{AC}\wedge \alpha^{DB}+a^{AC}\wedge a^{DB})\eta_{CD}\\&=&\underbrace{d_ta^{AB}+^{(3)}\!d\alpha^{AB}+(\alpha^{AC}\wedge a^{DB}+a^{AC}\wedge \alpha^{DB})\eta_{CD}}_{f_t^{\ AB}}+\underbrace{^{(3)}\!da^{AB}+a^{AC}\wedge a^{DB}\eta_{CD}}_{f^{AB}}\label{curv3}\\
&=&f_t^{\ AB}+f^{AB}
\end{eqnarray}
Therefore, for some `internal four form' $y_{ABCD}$ we have:

\begin{eqnarray}
y_{ABCD}F^{AB}\wedge F^{CD} &=& y_{ABCD}\left(f_t^{\ AB}+f^{AB}\right)\wedge\left(f_t^{\ CD}+f^{CD}\right)\\
     &=& y_{ABCD}\left(f_{t}^{AB}\wedge f^{CD} + f^{AB}\wedge f_{t}^{CD}\right) \\
        &=& 2y_{ABCD}f_{t}^{AB}\wedge f^{CD}
\end{eqnarray}
The action (\ref{ECMM2}) then becomes:

\begin{eqnarray}
 \kappa_{MM}\int  \epsilon_{ABCDE}V^{E}F^{AB}\wedge F^{CD}&=&  8\kappa_{MM}\int\phi \left(T_{[A}\varepsilon_{BCD]}   f_{t}^{AB}\wedge f^{CD} +S_{[A}\varepsilon_{BCD]} f_{t}^{AB}\wedge f^{CD}\right)\\
    &=&\kappa_{MM}\int \phi \left(2\epsilon_{IJKL} f_{t}^{IJ}\wedge f^{KL} +8S_{[A}\varepsilon_{BCD]} f_{t}^{AB}\wedge f^{CD}\right)\\
    &=&  4\kappa_{MM}\int\phi \varepsilon_{ijk}\left( \left(f^{4i}_{t}+f^{ti}_{t}\right)\wedge f^{jk} + f^{ij}_{t} \wedge \left(f^{4k}+f^{tk}\right)\right)\label{act1}
\end{eqnarray}
where $4T_{[A}\varepsilon_{BCD]} =\epsilon_{IJKL}\equiv \epsilon_{IJKL4}$ and recall that indices $I,J,K,L$ only go from $0$ to $3$.
Following from this, one may use the explicit form for the curvature contributions to (\ref{act1}) implied by (\ref{curv3})  along with the comparative decomposition given by (\ref{adecomp1}) to present the action in terms of conformal variables of (\ref{adecomp2}), yielding the following result:

 \begin{eqnarray}
 \label{cc1}
{\cal S}_{CC} &=& 4\kappa_{MM}\int \phi\varepsilon_{ijk} \left({\cal K}^{i}\wedge\bar{R}^{jk}+{\cal K}^{ij}\wedge \bar{T}^{k} \right)
 \end{eqnarray}
 where
 
 \begin{eqnarray}
\label{of1} {\cal K}^{i} &\equiv&  d_{t} E^{i}+^{(3)}d{\cal E}^{i}+{\cal E}^{l}\wedge \omega_{l}^{\phantom{j}i}+E^{l}\wedge \Omega_{l}^{\phantom{j}i}+{\cal C}\wedge E^{i}+C\wedge {\cal E}^{i}\\
\bar{T}^{i}&\equiv&  ^{(3)}d E^{i} + E^{l}\wedge \omega_{l}^{\phantom{l}i}+C\wedge E^{i}\\
{\cal K}^{ji} &\equiv& f^{ji}_{t} = d_{t} \omega^{ji} +^{(3)}d\Omega^{ji}+ \Omega^{jl}\wedge \omega_{l}^{\phantom{l}i} + \omega^{jl}\wedge \Omega_{l}^{\phantom{l}i}-{\cal F}^{[j}\wedge E^{i]}-F^{[j}\wedge {\cal E}^{i]}\\
\label{of2} \bar{R}^{ji} &\equiv & f^{ji} =  ^{(3)}d \omega^{ji} + \omega^{jl}\wedge \omega_{l}^{\phantom{l}i}-F^{[j}\wedge E^{i]}
 \end{eqnarray}
 
 We will refer to the action (\ref{cc1}) as \emph{Cartan Conformal Gravity}, stemming as it does from a connection $A^{AB}$ which includes, collectively, generators of translations on a model space as well as `point of contact' preserving rotation and conformal transformations. The utility of the null norm condition $|V|^{2}=0$ is to allow separation in the connection of the the `triad' field $\mathbf{E}^{i}$ from the gauge fields of the rotation and conformal transformations. In General Relativity plus a positive cosmological constant, the model space was de Sitter space. For Cartan Conformal Gravity, the model space is in fact the set of conformally flat three dimensional spaces, as shall be indicated by the solution presented in Section \ref{seceq}.
 
 For this theory, the only fields appearing with time derivatives are the fields $E^{i}$ and $\omega^{ji}$. The conformal gauge fields
$\mathbf{C}$ and $\mathbf{F}^{i}$ appear algebraically in the action. Although the presence of these auxiliary fields may seem unusual, it should be remembered that in the Palatini action the co-tetrad $e^{I}$ itself appears entirely without derivatives.

 The fields $\bar{T}^{i}$ and $\bar{R}^{ji}$ are suggestively named as they correspond respectively to the torsion and Riemannian
 curvature with corrective terms in conformal gauge fields $C$ and $F^{j}$ which ensure the required transformation properties
 under conformal transformations so as to preserve the action's invariance under them. However, it should be noted that the triad $E^{i}$ is \emph{not} identified simply with spatial parts of the co-tetrad $e^{i}$; rather it is proportional to a sum of $e^{i}$ and Lorentz boost fields $b^{i}$.
 
 Finally, it may be checked using the results of Appendix \ref{ctrans} that the functional form of the action (\ref{cc1}) is, as expected, invariant under dilation and special conformal transformations as well as three dimensional rotations.

\section{Equations of motion of  Cartan Conformal Gravity} 
\label{seceq}

We may recover the equations of motion for Cartan Conformal Gravity by the standard method of requiring stationarity of the action (\ref{cc1}) under small variations of fields. Note that in the construction of ${\cal S}_{CC}$ we have assumed that the null condition is satisfied at the level of the action, and so the stationary property of the action must apply also with respect to small variations of $\phi$ when the equations of motion are satisfied.

Variation with respect to ${\cal E}^{i}$, $\Omega^{ij}$, ${\cal F}^{i}$, ${\cal C}$, $F^{j}$, $C$, $\omega^{mn}$, $E^{m}$, and $\phi$ respectively yields:

\begin{eqnarray}
\label{cceqm1}0 &=& \varepsilon_{ijk}(C\wedge \bar{R}^{jk}+F^{j}\wedge \bar{T}^{k})
- \varepsilon_{mjk} \omega_{i}^{\phantom{i}m}\wedge \bar{R}^{jk}-\varepsilon_{ijk}\frac{^{3}d\left(\phi \bar{R}^{jk}\right)}{\phi} \\
0 &=& \varepsilon_{jmk}\left(E_{i}\wedge \bar{R}^{mk}-\omega^{m}_{\phantom{m}i}\wedge \bar{T}^{k}\right)
-\varepsilon_{imk}\omega_{j}^{\phantom{j}m}\wedge\bar{T}^{k} - \varepsilon_{ijk}\frac{^{3}d\left(\phi\bar{T}^{k}\right)}{\phi} \\
0&=&\varepsilon_{ijk}E^{j}\wedge \bar{T}^{k}\\
0&=&\varepsilon_{ijk} E^{i}\wedge \bar{R}^{jk}\\
0 &=& \varepsilon_{imk}{\cal K}^{i}\wedge E^{k}+\varepsilon_{mjk}{\cal E}^{j}\wedge \bar{T}^{k}\\
0&=& \epsilon_{ijk} {\cal K}^{ij} \wedge E^{k}+\varepsilon_{ijk}{\cal E}^{i}\wedge \bar{R}^{jk} \\
0 &=&-\varepsilon_{mnk} \frac{d_{t}\left(\phi\bar{T}^{k}\right)}{\phi}-\varepsilon_{imn}\frac{^{3}d\left(\phi {\cal K}^{i}\right)}{\phi} -
   \varepsilon_{imk}{\cal K}^{i}\wedge \omega_{n}^{\phantom{n}k} +\varepsilon_{ijn} {\cal K}^{i}\wedge \omega^{j}_{\phantom{j}m} +\varepsilon_{ink}\Omega^{i}_{\phantom{i}m}\wedge \bar{T}^{k} \nonumber\\
    &&
    -\varepsilon_{mjk}\Omega_{n}^{\phantom{n}j}\wedge \bar{T}^{k} + \varepsilon_{ijn} {\cal K}^{ij}\wedge E_{m}+ \varepsilon_{njk} {\cal E}_{m}\wedge \bar{R}^{jk}\\
 0 &=& \varepsilon_{mjk}\frac{d_{t}\left(\phi\bar{R}^{jk}\right)}{\phi}- \varepsilon_{ijm}\frac{^{3}d\left(\phi {\cal K}^{ij}\right)}{\phi}- \varepsilon_{ijk}\Omega_{m}^{\phantom{m}i}\wedge \bar{R}^{jk}
+\varepsilon_{mjk}{\cal C}\wedge \bar{R}^{jk}-\varepsilon_{ikm} {\cal K}^{i}\wedge F^{j} \nonumber \\
& &  -\varepsilon_{imk} {\cal F}^{i}\wedge \bar{T}^{k} 
-\varepsilon_{ijk} {\cal K}^{ij} \wedge \omega_{m}^{\phantom{m}k} +\varepsilon_{ijm}{\cal K}^{ij}\wedge C\\
\label{cceqm2}0 &=& \varepsilon_{ijk} \left({\cal K}^{i}\wedge\bar{R}^{jk}+{\cal K}^{ij}\wedge \bar{T}^{k}\right)
\end{eqnarray}

\subsection{A simple solution}

By inspection a particular solution to (\ref{cceqm1})-(\ref{cceqm2}) is:

\begin{eqnarray}
 \mathbf{E}^{i}_{\mu} &=& (0, \delta^{i}_{m})\qquad \mathbf{F}^{i}_{\mu} = (0,0) \qquad \mathbf{C}_{\mu} = (0,0) \qquad
 \mathbf{W}^{ik} = (0,0) \qquad \phi = 1 
 \end{eqnarray}
which implies that 

\begin{eqnarray}
\bar{R}^{jk} &=& 0 \qquad \bar{T}^{j}  =0
\qquad {\cal K}^{i} = 0 \qquad {\cal K}^{ij} = 0  \label{solit}
\end{eqnarray}

An analogous situation in the Macdowell-Mansouri case is the solution $F^{IJ}=0$ which corresponds to $R^{IJ}= \frac{1}{l^{2}}e^{I}\wedge e^{J}$ i.e. the Macdowell-Mansouri prescription singles out de Sitter space as a particular simple solution where the actual spacetime is identical to the model spacetime. Similarly in Cartan Conformal Gravity, the equations of motion allow a  simple solution wherein the `triad' field $E^{i}_{m}$ corresponds to that of three dimensional flat Euclidean space up to a space and time dependent conformal factor.
As may be expected in a theory violating Lorentz invariance, the picture of the gravitational field is one of spatial rather than spacetime structure. A suggestive indicator as to the generality of this behaviour comes via the $SO(1,4)$ and $C(3)$ invariant tensor $\mathfrak{g}_{\mu\nu}$, defined as follows:

\begin{eqnarray}
\mathfrak{g}_{\mu\nu} &\equiv & l^{2}\eta_{AB}D_{\mu}V^{A}D_{\nu}V^{B}  
\end{eqnarray}
where $D_{\mu}V^{A}\equiv \partial_{\mu}V^{A}-i{\cal A}^{A}_{\phantom{A}B}V^{B}$ is the covariant derivative of $V^{A}$.

For the case where $|V|^{2}=1$ we have that:

\begin{eqnarray}
\mathfrak{g}^{(MM)}_{\mu\nu} &=& \eta_{IJ}e^{I}_{\mu}e^{J}_{\nu}
\end{eqnarray}
i.e. in this Macdowell-Mansouri case, $\mathfrak{g}_{\mu\nu}$ reduces to the familiar spacetime metric tensor $g_{\mu\nu}$.
In the case of Cartan Conformal Gravity, we have that:

\begin{eqnarray}
\mathfrak{g}^{(CC)}_{\mu\nu} &= &\phi^{2} \delta_{ij}\left(\mathbf{E}^{i}\right)_{\mu}\left(\mathbf{E}^{j}\right)_{\nu}
\end{eqnarray}

Therefore in the preferred frame, corresponding to ${\cal E}^{i}=0$, the `spacetime metric' $\mathfrak{g}_{\mu\nu}$ has signature $(0,+,+,+)$.
Interestingly, the presence of a degenerate spacetime metric has much in common with geometric formulations of nonrelativistic gravitational theories, of which the Newton-Cartan theory is an example \cite{Andringa:2010it,Duval:2009vt,Duval:2011mi,Julia:1994bs}.
The ultimate role of a field such as $\mathfrak{g}_{\mu\nu}$ depends on the manner in which matter is coupled to $A^{AB}$ and $V^{A}$.
We now discuss in slightly more detail the scope for coupling to matter and providing dynamics for the field $V^{A}$.

\section{The dynamics of $V^{A}$}

One may expect $V^{A}$ to be described by an action with `kinetic term' and some accompanying mechanism to preferentially drive the quantity $|V|^{2}$ to a particular vacuum expectation value. One may then seek to construct an $SO(1,4)$ invariant action from $V^{A}$, and $D V^{A}$ as well as  the covariant `internal tensors' $\eta_{AB}$, $\epsilon_{ABCDE}$. The simplest possibility is:

\begin{eqnarray}
\label{sv1}
{\cal S}_{V1}  &=& \int \epsilon_{ABCDE}V^{E} DV^{A}\wedge D V^{B}\wedge D V^{C} \wedge D V^{D}
\end{eqnarray}

To get a sense of the `dynamics' implied by this action, we first consider the case where $V^{A}$ is space-like. One may assume a constant `background' value for $V^{A}$ and consider 
perturbations $v^{A}(x^{\mu})$ around this value for fixed $A^{C}_{\phantom{C}D}$. The perturbations are taken to be small in the sense that we expand the action (\ref{sv1}) to at-most quadratic order in $v^{A}$.
We choose to work in a gauge where $V^{A}=(1+v^{4})\delta^{A}_{4}$, which is always accessible if $V^{A}$ is space-like. Recalling the explicit form for the generators of $SO(1,4)$ we have that

\begin{eqnarray}
\label{cotedv}
D_{\mu}V_{(MM)}^{A} 
      & =& -(1+v^{4})e_{\mu}^{A}+ \partial_{\mu}v^{A}
\end{eqnarray}

Therefore we have that:

\begin{eqnarray}
\label{sv2}
{\cal S}_{V1(MM)}  &\propto
& \int \epsilon_{ABCD4}(1+v^{4})e^{A}\wedge e^{B}\wedge e^{C}\wedge e^{D}   \quad !
\end{eqnarray}

The underlying reason for this simple result is the fact that $\epsilon_{ABCDE}V^{E}$ is forced to be proportional to $\epsilon_{ABCD4}$; contractions over $A,B,C,D$ are then necessarily  over indices 0..3 which, by the gauge choice, involve no derivatives of $v^{A}$.
Therefore, the action (\ref{sv1}) in the regime of space-like $V^{A}$ seems not adequate to describe kinetic terms of $V^{4}$ nor 
a `potential' that might lend itself towards a non-vanishing vacuum expectation value for $|V|^{2}$. Rather, the combined action (\ref{ECMM2}) and (\ref{sv2})  is algebraic and linear in $v^{4}$, and so essentially acts as a Lagrange multiplier, enforcing a constraint between $\epsilon_{IJKL}F^{IJ}\wedge F^{KL}$ of (\ref{ECMM2}) and terms of $\epsilon_{IJKL}e^{I}\wedge e^{J} \wedge e^{K}\wedge e^{L}$ of (\ref{sv2}). The field $v^{4}$ itself would be recoverable from the equations $\delta({\cal S}_{V1}+{\cal S}_{MMV})/\delta A^{ab}$. 
The presence of the new constraint and possible variation of $v^{4}$ away from $0$ makes the similarity of ${\cal S}_{V1}+{\cal S}_{MMV}$ to General Relativity unclear.

We note that the covariant derivative's ability to approximate the co-tetrad around certain backgrounds, as evident in (\ref{cotedv}) and
(\ref{sv2}), makes it useful for constructing realistic matter Lagrangians in an $SO(1,4)$ invariant manner \cite{WestmanZlosnik2012}. For example, it may be shown that a Lagrangian of the following form corresponds to a Klein-Gordon Lagrangian for appropriate  $\kappa_{ABCDE}$ (assumed to be formed only from combinations of $\epsilon_{ABCDE}$, $\eta_{AB}$, $V^{A}$) and 
$D_{\mu}V_{(MM)}^{A}    \sim-e_{\mu}^{A}$:
\begin{eqnarray}
DV^{A}\wedge DV^{B}\wedge DV^{C} \wedge\left(Y^{E}\kappa_{ABCDE}\wedge DY^{D}\right) &\propto& e^{I} \wedge e^{J} \wedge e^{K} \wedge \left(Y^{E}\kappa_{IJKDE} DY^{D}\right) \label{kgact}
\end{eqnarray}
The component $Y^{4}$ ultimately plays the role of a Klein-Gordon scalar field, whilst $Y_{J}$ is representative of its spacetime derivatives according to the co-tetrad (itself defined by appropriate behaviour of $V^{A}$). Note that the action (\ref{kgact}) requires no basic counterpart to the inverse-tetrad field $e^{\mu}_{J}$ at the level of the action.
Actions such as (\ref{kgact}) are  reminiscent of so-called first order DKP formulations 
of spin-0 and spin-1 fields \cite{Lunardi:1999jq,Casana:2002fu,Pavlov:2006kh}. For the case that $Y^{A}$ happens to  be $V^{A}$ itself, the only non-trivial possibility for $\kappa_{ABCDE}$ is $\epsilon_{ABCDE}$ \cite{WestmanZlosnik2012} which, as we have seen, is in several senses unsatisfactory. See \cite{Wilczek:1998ea} for an alternative approach to accommodating the $V^{A}$ field and matter in Macdowell-Mansouri gravity via the introduction of a preferred volume element.

For the case of Cartan Conformal Gravity we have

\begin{eqnarray}
\label{cotedv}
D_{\mu}V_{(CC)}^{A} 
      & =& -\phi\mathbf{E}^{A}_{\mu} -(\mathbf{C}_{\mu}-\frac{1}{\phi}\partial_{\mu}\phi)V^{A}
\end{eqnarray}

From this it may be checked that in the Cartan Conformal Gravity case, the integral (\ref{sv1}) is identically zero, the reason being that it now contains a contraction $\epsilon_{ABCDE}V^{D}V^{E}$.
The action ${\cal S}_{V1}$ does not seem suitable for prescribing dynamics of $V^{A}$, and it seems difficult to envision a suitable action built from $A^{AB}$ and $V^{A}$. If these fields cannot collectively define a realistic dynamics amongst themselves other than by simply constraining the desired norm $|V|^{2}$ via the use of a Lagrange multiplier \cite{Stelle:1979va}, then it would seem necessary to introduce more complicated structure into the theory. 

Thusfar we have considered gravitation from the perspective of an $SO(1,4)$ (or $C(3)$) gauge field and a field $V^{A}$ valued in its matrix representation. However, the difficulty with constructing dynamics of $V^{A}$ may suggest that an alternative is called for. One possibility is that it is more appropriate to regard gravity as belonging to the four dimensional representation of $Spin(1,4)$ (the double cover of $SO(1,4)$, and sometimes referred to as its `Spin-1/2' representation), for which the explicit form for the generators $S_{AB}$ is given in Appendix \ref{s12}. It may be shown that an action giving equivalent physics to (\ref{mmm1}) is given by \cite{Randono:2010cq}:

\begin{eqnarray}
\label{s123}
{\cal S}_{MM(SP)} &=& \kappa_{MM(SP)}\int Tr\left( \gamma^{5} F\wedge F\right)
\end{eqnarray}

It has been shown by Randono \cite{Randono:2010ym} that a spinor multiplet with suitably constrained invariants can play the role of the $\gamma^{5}$ matrix in (\ref{s123}) and allow for more general gravitational actions than (\ref{mmm1}). Moreover it may be checked that the Lorentz violating case of Cartan Conformal Gravity may presumably by a configuration of fields reducing also by tensor product to the Dirac matrix $\gamma^{0}$.
 However, as in the case of the field $V^{A}$, it is not yet clear how to construct suitable kinetic terms for these fields.
 
 As discussed in \cite{Wilczek:1998ea}, it is possible that the internal group could be substantially larger than $SO(1,4)$. Conceivably
 a counterpart of $V^{A}$ valued in this larger group may be able, along with the larger groups connection coefficient one-forms, to account for their own dynamics.

\section{Discussion}

The isomorphism between the group $SO(1,4)$ and the conformal group $C(3)$ has been seen to be useful in the interpretation of the behaviour of the gravitational action (\ref{ECMM2}) for the case where $|V|^{2}$ is null.
The resulting theory, termed Cartan Conformal Gravity, has been seen to be more readily interpreted in terms of spatial rather than spacetime structure, it's field equations describing the evolution of what may be termed conformal spatial Riemmanian curvature and torsion. However, the spatial triad $E^{i}$ involved in these quantities is not simply the spatial spacetime co-tetrad $e^{i}$ but is rather a sum of this field with the spacetime-boost field $b^{i}$, which in a more familiar setting corresponds to the extrinsic curvature of constant time slices in the Palatini formulation of General Relativity; this points towards the possibility that other degrees of freedom such as 
$V^{A}$ in the gravitational sector may lead to very different relationships between geometry and the gravitational field in different regimes.
Moreover, the interpretation of Cartan Conformal Gravity in terms of conformal space-like geometry is entirely as a result of the choice of a positive cosmological constant. The appropriate group for General Relativity plus a negative cosmological constant is $SO(2,3)$, which is isomorphic to $C(1,2)$ i.e. the group of transformations conformally preserving the Minkowski metric of signature $(-,+,+)$, suggesting that the counterpart Cartan Conformal Gravity would take the form a theory of spacetime structure but of a lower dimensionality and varying along a preferred space-like direction.

Returning to the $SO(1,4)$ case, one may yet choose to use $C(3)$ variables for the case where $|V|^{2}$ is a space-like constant i.e. General Relativity with a positive cosmological constant. The possible utility of such variables is the subject of ongoing investigation. In any case it is to be expected that  geometric interpretation of the gravitational field is more appropriately expressed in terms of $e^{J}$ in this case, much as it appears more appropriate to use $E^{i}$ in the Cartan Conformal Gravity case. However, we note that interpretation of the theory of General Relativity, with or without cosmological constant, in terms of spatial conformal invariance (though the interpretation of the term differs from that used in this article) has been a subject of much recent interest \cite{Gomes:2011au,Gomes:2011dc,Gomes:2010fh,Barbour:2010xk,Barbour:2011dn,Westman:2009bh} and it may be useful to investigate possible parallels between approaches.

Due to the difficulty in constructing dynamics for the field $V^{A}$ (or indeed whatever field may play a similar role to $V^{A}$), the general consequences of Cartan Conformal Gravity are as yet unclear, though the form of the action (\ref{cc1}) suggests something markedly different to General Relativity. It would seem that a necessary next step then is to see whether an alternative to $V^{A}$ may be found.
It should be emphasized that not only is it desirable to construct dynamics for the role here played by $V^{A}$, it is necessary also to cast familiar matter fields in terms of actions invariant under a group larger than $SO(1,3)$. Though in the case of constant space-like $|V|^{2}$ it appears possible to recover familiar classical equations of motion for matter \cite{WestmanZlosnik2012}, much as one can recover General Relativity, it is not clear if there are differences in the behaviour of matter beyond this. Of particular interested is the relationship between the scale $l$ and the measured value of the cosmological constant.

Finally, recall that Cartan Conformal Gravity follows from the case $V^{0}=-V^{4}$. Is the alternative possibility $V^{0}=V^{4}$ a physically different theory? By inspection it would appear that the resulting theory is identical, so long as one allows for the roles of
 $\mathbf{F}^{i}$ and $\mathbf{E}^{i}$ to be interchanged, and the attribution of a different conformal weight for the field $\phi$ and the new `triad'  $\mathbf{F}^{i}$. This equivalence would appear to hold even though the interpretation of invariance properties under internal transformations on the model space differs significantly. The reason for this is the similar manner in which the generators  of translations $p_{m}$ and special conformal translations $k_{m}$ appear in the Lie algrebra of the group $C(3)$. Notably this ambiguity in the relationship between gravitational fields and particular connection coefficients is present in attempts to construct spacetime gauge theories for the four dimensional conformal group $C(4)$ \cite{Wheeler:1991ff}. \newline \newline
\textbf{Acknowledgements}: We would like to thank Florian Girelli and Tim Koslowski for useful discussions. This research was supported by the Perimeter Institute-Australia Foundations (PIAF) program. 

\appendix
\section{Representations of $SO(1,4)$}
The group $SO(1,4)$ consists of all linear transformations that preserve the line element $\eta_{AB}=diag(-1,1,1,1,1)$. The Lie algebra for $SO(1,4)$ is:
\begin{eqnarray}
\label{commu}
\left[S_{AB},S_{CD}\right]= -i\left(\eta_{AC}S_{BD}-\eta_{AD}S_{BC}-\eta_{BC}S_{AD}+\eta_{BD}S_{AC}\right).
\end{eqnarray}

A field $\psi$ in a particular linear representation of $SO(1,4)$ transforms under an infinitesimal gauge transformation as $\tilde{\psi}\rightarrow U\psi$, where $U$ is defined as follows:

\begin{eqnarray}
\tilde{\psi}^{\Omega} = (e^{-i\Theta^{AB}S_{AB}})^{\Omega}_{\phantom{\Omega}\Gamma}\psi^{\Gamma}
\end{eqnarray}

Given this, one may define the covariant derivative $D_{\mu}\psi^{\Omega}= \partial_{\mu}\psi^{\Omega}- i {\cal A}^{\phantom{\mu}\Omega}_{\mu\phantom{\mu}\Gamma}\psi^{\Gamma}$ which transforms as $\tilde{D}_{\mu}\tilde{\psi}^{\Omega} \rightarrow  (e^{-i\Theta^{AB}S_{AB}})^{\Omega}_{\phantom{\Omega}\Gamma}D_{\mu}\psi^{\Gamma}$ (i.e. it transforms \emph{homogeneously}) as long as the \emph{connection} spacetime one-form ${\cal A}^{\phantom{\mu}\Omega}_{\mu\phantom{\mu }\Gamma}$ transforms as:

\begin{eqnarray}
\label{atrans}
\tilde{{\cal A}} = U{\cal A} U^{-1} - i \left(dU\right)U^{-1}
\end{eqnarray}

The abstract `indices'  $\Omega,\Gamma$ refer to a general representation of $SO(1,4)$. Of particular use will be the so-called Spin-1 and Spin-$1/2$ representations of the group which will be briefly discussed as follows.

\subsection{Spin-1 representation of $SO(1,4)$}
In this representation, the generators $S_{AB}$ take the form of 5-dimensional matrices. In index notation the generators take the following form:

\begin{eqnarray}
\label{sp1reps}
(S_{AB})^{C}_{\phantom{C}D} =- i\left(\eta_{AD}\delta_{B}^{\phantom{B}C}-\eta_{BD}\delta_{A}^{\phantom{A}C}\right)
\end{eqnarray}

Accordingly we may define the connection ${\cal A}^{\phantom{\mu}C}_{\mu\phantom{C} D}$ in terms of the generators $(S_{ab})^{C}_{\phantom{C}D} $ and 
`connection coefficient' one-form fields $A^{\phantom{\mu}AB}_{\mu}$ as follows:

\begin{eqnarray}
\label{aexp}
{\cal A}^{\phantom{\mu}C}_{\mu\phantom{C} D}=  \frac{1}{2} A^{AB}_{\mu}(S_{AB})^{C}_{\phantom{C}D}
\end{eqnarray}

Note the spin-1/matrix representation the generator labels and matrix component indices may be used interchangeably. Indices may be raised and lowered with $\eta^{AB}$ and $\eta_{AB}$.

We may in turn define the curvature ${\frak F}^{\phantom{\mu\nu}AB}_{\mu\nu}$:

\begin{eqnarray}
{\frak F}^{\phantom{\mu\nu}AB}_{\mu\nu} = \left(d{\cal A}^{AB}\right)_{\mu\nu} - i [{\cal A}^{A}_{\phantom{A}C},{\cal A}^{CB}]_{\mu\nu}
\end{eqnarray}
Given the transformation rule for (\ref{atrans}), it is easily checked that ${\frak F}^{\phantom{\mu\nu}AB}_{\mu\nu}$ transforms homogeneously under gauge transformations.

As for the case of ${\cal A}^{\phantom{\mu}C}_{\mu\phantom{\mu C} D}$ we may define representation independent fields $F^{\phantom{\mu\nu}AB}_{\mu\nu}$ as follows:

\begin{eqnarray}
{\frak F}^{\phantom{\mu\nu}C}_{\mu\nu \phantom{\mu\nu}D} =  \frac{1}{2} F^{\phantom{\mu\nu}AB}_{\mu\nu}(S_{AB})^{C}_{\phantom{C}D}
\end{eqnarray}

It is readily checked that:

\begin{eqnarray}
F^{ab}_{\mu\nu} = \left(dA^{AB}\right)_{\mu\nu}+ \left(A^{A}_{\phantom{a}C}\wedge A^{CB}\right)_{\mu\nu}
\end{eqnarray}

\subsection{Spin-1/2 representation of $SO(1,4)$}
\label{s12}
In this representation, the generators $S_{AB}$ take the form of 4-dimensional matrices. We choose the following explicit form for the generators $S_{AB}$:

\begin{eqnarray}
S_{AB} =- \frac{i}{4}\left[\Gamma_{A},\Gamma_{B}\right]
\end{eqnarray}

where $\Gamma_{A}= (\gamma_{I},i\gamma_{5})$ where $\gamma_{I}$ are the Dirac gamma matrices and $\gamma_{5}\equiv \left(\frac{i}{4!}\right)\epsilon_{IJKL}\gamma^{I}\gamma^{J}\gamma^{K}\gamma^{L}$. It may readily be checked that the matrices $\Gamma_{A}$ satisfy the Clifford algebra:

\begin{eqnarray}
\Gamma_{A}\Gamma_{B}+\Gamma_{B}\Gamma_{A}= -2\eta_{AB}
\end{eqnarray}

\section{Gauge transformations}
\label{ctrans}

The Lie algebra (\ref{commu}) of $SO(1,4)$ and $C(3)$ may be written more compactly in the form

\begin{eqnarray}
[S_{AB},S_{CD}]=  f_{ABCD}^{\phantom{ABCD}EF}S_{EF}
\end{eqnarray}
where

\begin{eqnarray}
f^{AB}_{\phantom{AB}CDEF} =- \delta^{A}_{C}\delta^{B}_{E}\eta_{FD}+\delta^{A}_{D}\delta^{B}_{E}\eta_{FC}+\delta^{B}_{C}\delta^{A}_{E}\eta_{FD}-\delta^{B}_{D}\delta^{A}_{E}\eta_{FC}
\end{eqnarray}
From this it follows that under an infinitesimal gauge transformation, the connection coefficient fields $A^{AB}$  transform as follows:

\begin{eqnarray}
A^{AB} \rightarrow A^{AB}+ d\Theta^{AB} - \frac{1}{2}f^{AB}_{\phantom{AB}CDEF}\Theta^{CD}A^{EF}
\end{eqnarray}

We now examine in detail how the fields $A^{AB}$ transform under the symmetries displayed by the action (\ref{cc1}).

\subsection{Change under special conformal transformations}
\label{sct}

Recalling the isomorphism between $SO(1,4)$ and $C(3)$, we see that an infinitesimal special conformal transformation is generated by a combined `spatial co-tetrad translation' and boost:  $k_{i} = S_{4i}-S_{0i}$ and can be parameterized by three numbers $\Theta^{i}_{(K)}$ where

\begin{eqnarray}
\Theta^{CD} &=& 2\left(\delta^{[C}_{4}\delta^{D]}_{i}-\delta^{[C}_{0}\delta^{D]}_{i}\right)\Theta_{(K)i}
\end{eqnarray}
Therefore we have that

\begin{eqnarray}
A^{AB}& \rightarrow& A^{AB}+2\left(\delta^{[A}_{4}\delta^{B]}_{i} -\delta^{[A}_{0}\delta^{B]}_{i}\right) d\Theta_{(K)}^{i} - \left(f^{AB}_{\phantom{AB}4iEF}- f^{AB}_{\phantom{AB}0iEF}\right)\Theta_{(K)}^{i}A^{EF} 
\end{eqnarray}
and from this it follows that

\begin{eqnarray}
 \mathbf{F}^{j} &\rightarrow &  \mathbf{F}^{j}+2d\Theta_{(K)}^{j}  +2\Theta^{i}_{(K)}\mathbf{W}^{j}_{\phantom{j}i}+2\Theta^{j}_{(K)}\mathbf{C}\\
 \mathbf{W}^{ij}& \rightarrow& \mathbf{W}^{ij} + 2\Theta^{[i}_{(K)}\mathbf{E}^{j]} \\
 \mathbf{C} &\rightarrow & \mathbf{C} -\mathbf{E}^{i}\Theta_{i(K)}\\
 \mathbf{E}^{j} &\rightarrow &\mathbf{E}^{j} 
 \end{eqnarray}

  \begin{eqnarray}
  {\cal K}^{i}& \rightarrow & {\cal K}^{i}  \qquad  \bar{T}^{i}\rightarrow  \bar{T}^{i} \\
   {\cal K}^{ij} &\rightarrow &  -2\Theta_{(K)}^{[i} {\cal K}^{j]} \qquad
   \bar{R}^{ij} \rightarrow   -2\Theta_{(K)}^{[i}\bar{T}^{j]}
    \end{eqnarray}

 \subsection{Change under dilations}
 \label{dct}
 
 A dilation may be parameterized by the following choice:
 
 \begin{eqnarray}
 \Theta^{CD}= 2 \delta_{0}^{[C}\delta^{D]}_{4}\Theta_{(D)}
 \end{eqnarray}
Therefore under a dilation we have:

\begin{eqnarray}
A^{AB} \rightarrow A^{AB}+2\delta^{[A}_{0}\delta^{B]}_{4} d\Theta_{(D)} - f^{AB}_{\phantom{AB}04EF}\Theta_{(D)}A^{EF}
\end{eqnarray}
 and hence
 
\begin{eqnarray}
 \mathbf{F}^{j} &\rightarrow & \left(1-\Theta_{(D)}\right) \mathbf{F}^{j}\\
 \mathbf{W}^{ij}& \rightarrow& \mathbf{W}^{ij}  \\
 \mathbf{C} &\rightarrow & \mathbf{C} -d\Theta_{(D)}\\
 \mathbf{E}^{j} &\rightarrow &\left(1+\Theta_{(D)}\right)\mathbf{E}^{j} \\
 \phi &\rightarrow & \left(1-\Theta_{(D)}\right)\phi
 \end{eqnarray}

 \begin{eqnarray}
{\cal K}^{i}& \rightarrow & \left(1+\Theta_{(D)}\right) {\cal K}^{i} \qquad {\bar T}^{i} \rightarrow (1+\Theta_{(D)}){\bar T}^{i} \\
   {\cal K}^{ij} &\rightarrow & {\cal K}^{ij} \qquad  \bar{R}^{ij} \rightarrow   \bar{R}^{ij}
    \end{eqnarray}
 
 \subsection{Change under Rotations}
 
 We parameterize a rotation by the following choice:
 
 \begin{eqnarray}
 \Theta^{CD}= 2 \delta_{i}^{[C}\delta^{D]}_{j}\Theta^{ij}_{(R)}
 \end{eqnarray}

where $\Theta_{ij(R)}=-\Theta_{ji(R)}$ i.e. there are three independent possible spatial rotations.

\begin{eqnarray}
A^{AB} \rightarrow A^{AB}+2 \delta_{i}^{[A}\delta^{B]}_{j}d\Theta^{ij}_{(R)}- f^{AB}_{\phantom{AB}ijEF}\Theta^{ij}_{(R)}A^{EF}
\end{eqnarray}

Therefore we have

\begin{eqnarray}
\mathbf{F}^{k} &\rightarrow & \mathbf{F}^{k} - 2\Theta^{k}_{(R)j}\mathbf{F}^{j}\\
\mathbf{W}^{ij} &\rightarrow &  2 d \Theta_{(R)}^{ij} - 2\Theta^{i}_{(R)k}\mathbf{W}^{kj}
- 2\Theta^{j}_{(R)k}\mathbf{W}^{ik}\\
\mathbf{C} &\rightarrow & \mathbf{C} \\
\mathbf{E}^{k} &\rightarrow& \mathbf{E}^{k} -2\Theta^{k}_{(R)j}\mathbf{E}^{j}
\end{eqnarray}

\subsection{Change under diffeomorphisms}

Finally for a spacetime diffeomorphism generated by a vector field $\xi^{\mu}$ we have:

\begin{eqnarray}
(\mathbf{F}^{k})_{\mu} &\rightarrow& (\mathbf{F}^{k})_{\mu} +  \xi^{\alpha}\partial_{\alpha}(\mathbf{F}^{k})_{\mu}+(\mathbf{F}^{k})_{\alpha}
\partial_{\mu}\xi^{\alpha}\\
(\mathbf{W}^{ij})_{\mu} &\rightarrow& (\mathbf{W}^{ij})_{\mu} +  \xi^{\alpha}\partial_{\alpha}(\mathbf{W}^{ij})_{\mu}+(\mathbf{W}^{ij})_{\alpha}
\partial_{\mu}\xi^{\alpha}\\
(\mathbf{C})_{\mu} &\rightarrow& (\mathbf{C})_{\mu} +  \xi^{\alpha}\partial_{\alpha}(\mathbf{C})_{\mu}+(\mathbf{C})_{\alpha}
\partial_{\mu}\xi^{\alpha}\\
(\mathbf{E}^{k})_{\mu} &\rightarrow& (\mathbf{E}^{k})_{\mu} +  \xi^{\alpha}\partial_{\alpha}(\mathbf{E}^{k})_{\mu}+(\mathbf{E}^{k})_{\alpha}
\partial_{\mu}\xi^{\alpha}
\end{eqnarray}

\section{Bibliography}
\bibliographystyle{hunsrt}
\bibliography{references}
\end{document}